\documentclass{IEEEtran}
\usepackage{cite}
\usepackage{amsmath,amssymb,amsfonts}
\usepackage{algorithmic}
\usepackage{graphicx}
\usepackage{textcomp}
\usepackage{threeparttable}
\usepackage{multicol, blindtext,xcolor}
\def\BibTeX{{\rm B\kern-.05em{\sc i\kern-.025em b}\kern-.08em
    T\kern-.1667em\lower.7ex\hbox{E}\kern-.125emX}}
\begin{document}
\title{Analytical Design of Printed-Circuit-Board (PCB) Metagratings for Perfect Anomalous Reflection}
\author{Oshri Rabinovich and Ariel Epstein, \IEEEmembership{Member, IEEE}
\thanks{The authors are with the Andrew and Erna Viterbi Faculty of Electrical Engineering, Technion - Israel Institute of Technology, Haifa 3200003, Israel (e-mail: oshrir@technion.ac.il; epsteina@ee.technion.ac.il).}
\thanks{Manuscript received XX,YY,2017; revised XX,YY, 2017.}}


\maketitle

\begin{abstract}
We present an analytical scheme for the design of realistic metagratings for wide-angle engineered reflection. These recently proposed planar structures can reflect an incident plane wave into a prescribed (generally non-specular) angle with very high efficiencies, using only a single meta-atom per period. Such devices offer a means to overcome the implementation difficulties associated with standard metasurfaces (consisting of closely-packed subwavelength meta-atoms) and the relatively low efficiencies of gradient metasurfaces. In contrast to previous work, in which accurate systematic design was limited to metagratings unrealistically suspended in free space, we derive herein a closed-form formalism allowing realization of printed-circuit-board (PCB) metagrating perfect reflectors, comprised of loaded conducting strips defined on standard metal-backed dielectric substrate. The derivation yields a detailed procedure for the determination of the substrate thickness and conductor geometry required to achieve unitary coupling efficiencies, without requiring \emph{even a single} full-wave simulation. Our methodology, verified via commercial solvers, ultimately allows one to proceed from a theoretical design to synthesis of a full physical structure, avoiding the time-consuming numerical \textcolor{black}{optimizations} typically involved in standard metasurface design.
\end{abstract}

\begin{IEEEkeywords}
Metagrating, anomalous reflection, wire grids
\end{IEEEkeywords}

\section{Introduction}
\label{Introduction}
\IEEEPARstart{R}{esearch} on beam-manipulating devices has experienced significant growth in the last few years, especially with the rapid developments in the field of metasurfaces.
Among these developments, it was shown that one can manipulate wavefronts using gradient metasurfaces following the generalized Snell's law \cite{yu2011light, sun2012gradient}, or by specifying the polarizability distribution needed for a surface to \textcolor{black}{implement} a prescribed field transformation using the associated generalized sheet transition conditions (GSTC) \cite{pfeiffer2013metamaterial,selvanayagam2013discontinuous, epstein2014passive,pfeiffer2014high,asadchy2015functional,achouri2015general,epstein2016huygens,epstein2016arbitrary,epstein2016synthesis,
asadchy2016perfect,estakhri2016wave}. While initial attempts to use gradient metasurfaces to deflect an incoming plane wave to a prescribed angle in transmission or reflection suffered from low efficiencies \cite{yu2011light}, accurate solutions based on bianisotropic GSTCs were later found to allow optimal coupling efficiencies \cite{epstein2016arbitrary,asadchy2016perfect,epstein2016synthesis}. However, \textcolor{black}{practical realization of the latter} still poses significant challenges, 
as it involves microscopic design of many (generally different) meta-atom structures per wavelength \textcolor{black}{to implement the discretized version of the GSTCs}, typically relying on time-consuming numerical simulations \cite{epstein2016huygens,epstein2016cavity,epstein2016arbitrary,pfeiffer2014bianisotropic}.

\begin{figure*}[t]
\centering
\centerline{\includegraphics[width=7in]{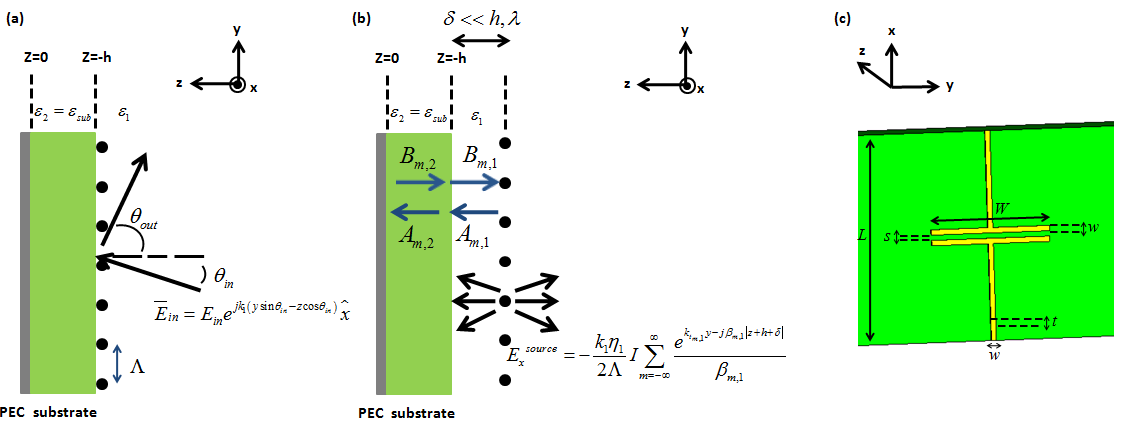}}
\caption{Physical configuration of the PCB-compatible metagrating. (a) A grid of loaded strips (with period $\Lambda$) in front of a dielectric substrate backed by a PEC layer. (b) An infinitesimal layer of thickness $\delta<<\lambda,h$ introduced in order to properly account for the field created by the grid. The field can be written as a sum of Hankel functions that can be interpreted as a sum of Floquet-Bloch modes using the Poisson formula (See Section \ref{subsec:Formulation}). (c) Unit cell configuration comprising a single meta-atom built of a printed capacitor. The capacitor width is $W$. The trace separation is $s$, its width is $w$ and the copper thickness is $t.$}
\label{metagrating}
\end{figure*}

A way of overcoming this obstacle was presented in \cite{ra2017meta}, where the concept of metagratings was rigorously introduced and systematically explored. It was demonstrated that using a single polarizable element per macro-period (comparable to the wavelength), one can manipulate the wavefront in order to get a reflection angle different than the specular with \textcolor{black}{unitary} efficiency. 
This is possible due to the fact that the period of a metagrating can be tuned such that only two propagating Floquet-Bloch (FB) modes are allowed (the specular and the anomalous reflection), while the rest are evanescent \cite{russell1986optics}. Thus, to obtain full coupling of the incoming beam into the desirable mode, one is merely required to eliminate the specular reflection, which can be achieved with a single meta-atom (single degree of freedom). In contrast,  metasurfaces are designed to realize a prescribed \emph{field transformation}, forbidding excitation of \emph{any} spurious FB mode, either propagating or evanescent \cite{epstein2016huygens,estakhri2016recent}. This strict requirement substantially increases the design complexity associated with metasurfaces, as it requires, in principle, control of an infinite number of degrees of freedom. 

It is important to note that the basic design principles of metagratings rely on the long-time research on diffraction gratings \cite{perry1995high,destouches200599}, where spatially-modulated dielectric structures were used to reduce undesirable scattering. However, the recent work by Ra'di \textit{et al.} \cite{ra2017meta} is revolutionary in the sense that it provides a means for rigorous semianalytical synthesis of the macro-period, replacing the intuitive approaches or numerical optimization methods that were usually carried out as the main design schemes for conventional diffraction gratings. 

Inspired by \textcolor{black}{these ideas}, several extensions of the analytical model developed in \cite{ra2017meta} have been presented more recently, enabling rigorous synthesis and analysis of metagratings based on electrically polarizable particles in the form of dielectric rods \cite{chalabi2017efficient} and capacitively loaded wires \cite{epstein2017unveiling}.
Nonetheless, all of these reports considered metagratings composed of meta-atoms unrealistically suspended in free space above a ground plane, thus not suitable for physical implementation. On the other hand, the experimental demonstrations of metagrating-based diffraction engineering that have been reported lately \cite{yang2017freeform,sell2017large,paniagua2017metalens,khaidarov2017asymmetric,wong2017perfect}, were designed based on full-wave numerical optimization of the meta-atoms, requiring substantial computational effort.

In this paper, we are filling this gap, presenting a complete analytical scheme for the design of realistic fabrication-ready metagratings for perfect anomalous reflection.
The devices are based on loaded metallic strips defined on a dielectric substrate,  backed by a metallic ground plane. Extending the analytical model of \cite{epstein2017unveiling} to rigorously account for the dielectric substrate, we formulate the physical conditions the metgarating configuration must fulfill in order to fully couple an incident plane wave to a given (non-specular) reflected mode. For a desired reflection angle, these conditions yield the substrate thickness and the capacitive load impedance of each strip required to guarantee unitary coupling efficiencies. Importantly, we show that the trace geometry corresponding to this capacitive loading can be assessed in closed-form, resulting in \textcolor{black}{complete} \emph{detailed} design specifications, without requiring even a \emph{single} full-wave simulation. As indicated by previous reports, these realistic perfect anomalous reflectors feature relatively large bandwidths and very small losses, which can be fully predicted by the detailed model and elucidated in correspondence to our previous observations \cite{epstein2017unveiling}.

These results, verified using \textcolor{black}{commercial} solvers, form an efficient synthesis procedure for metagratings, enabling a complete bottom-up design methodology connecting the analytical model with the final printed circuit board (PCB) layout, without any full-wave optimization whatsoever. The realistic model, rigorously incorporating the dielectric substrate, conductor losses, and frequency dependency, is expected to accelerate the (theoretical and experimental) development of these novel, simple yet powerful, devices for advanced diffraction engineering.

\section{Theory}
\label{sec:Theory}
\subsection{Formulation}
\label{subsec:Formulation}
We consider a 2D ($\partial/\partial x=0$) configuration comprised of loaded conducting strips positioned on a dielectric substrate with permittivity $\varepsilon_{2}=\varepsilon_\mathrm{sub}$ \textcolor{black}{and} thickness $h$, backed by a perfect-electric-conductor (PEC) layer at $z=0$ (Fig. \ref{metagrating}); this structure is surrounded by a medium with dielectric constatnt $\varepsilon_{1}=\varepsilon_\mathrm{air}$, occupying the half-space $z<-h$. The conducting strips are of width $w$ and thickness $t$ [Fig. \ref{metagrating}(c)], and are repeated to form a periodic structure in the plane $z=-h$, with $L\ll\lambda$ and $\Lambda$ being the periodicities along the $x$ and $y$ axes, respectively.
The conductor dimensions are assumed to satisfy $t\ll w\ll \Lambda,\lambda$; hence, throughout the paper, we use the flat wire model presented in \cite{tretyakov2003analytical} regarding the strips as small conducting cylinders with effective radius $r_\mathrm{eff}=w/4$. The strips are loaded by a lumped impedance [printed capacitor in Fig. \ref{metagrating}(c)]; due to the deep subwavelength $x$-periodicity $L\ll\lambda$, they effectively form a distributed impedance of $\widetilde Z$ per unit length along this dimension. 

We excite the structure with a transverse electric (TE) polarized plane wave ($E_{z}=E_{y}=H_{x}=0$) with an angle of incidence $\theta_\mathrm{in}$ with respect to the normal, as shown in Fig. \ref{metagrating}(a). The surrounding medium is indicated by the subscript 1, with wave number $k_{1}=\omega\sqrt{\mu_1\varepsilon_1}$ and wave impedance $\eta_{1}=\sqrt{\mu_1/\varepsilon_1}$, while the dielectric substrate will be indicated by the subscript 2, with wave number $k_{2}=\omega\sqrt{\mu_2\varepsilon_2}$ and wave impedance $\eta_{2}=\sqrt{\mu_2/\varepsilon_2}$; time harmonic dependency of $e^{j\omega t}$ is assumed and suppressed.


Our objective is to design a metagrating that reflects a plane wave with a given angle of incidence $\theta_\mathrm{in}$ to a prescribed output angle $\theta_\mathrm{out}$, with unitary efficiency. In order to do so, we seek to properly set the two degrees of freedom available in our configuration, namely, the thickness of the dielectric substrate, $h$, and the load impedance-per-unit-length, $\widetilde Z$.

We start by harnessing the superposition principle to separate the fields in the problem into two sets \cite{ra2017meta,epstein2017unveiling}: the first corresponds to the fields due to the incoming plane wave scattered off the structure in the absence of the loaded strips (external fields); the second set involves the fields generated by the grid of loaded strips due to the current $I$ induced on it by these external fields. Both scenarios are subject to the same boundary conditions, associated with the dielectric substrate and the back-PEC.

To find the external fields, $E_x^\mathrm{ext}$, we first write them as a superposition of forward and backward propagating plane waves in each of the media $1$, $2$. Subsequently, we derive the tangential magnetic fields according to Maxwell's equations, $H_{y}(y,z)=-\frac{1}{jk\eta}\frac{\partial}{\partial z} E_{x}(y,z)$, and impose the configuration's boundary conditions: continuity of the tangential electric and magnetic fields at the air-dielectric boundary, and vanishing of the tangential electric field at the PEC interface. For a given amplitude $E_\mathrm{in}$ of the incoming plane wave, the electric field everywhere in space \textcolor{black}{can} be resolved as
\begin{equation}
\label{eq:E_ext_air}
\begin{aligned}
E^\mathrm{ext}_{x,1}(y,z)=E_\mathrm{in}\left[e^{-j\beta_{0,1}z}+R_{0}e^{+j\beta_{0,1}(z+2h)}\right]e^{jk_{t_{0,1}}y}
\end{aligned}
\end{equation}
in medium 1 ($z<-h$), and
\begin{equation}
\label{eq:E_ext_dielectric}
\begin{aligned}
E^\mathrm{ext}_{x,2}(y,z)=-E_\mathrm{in}T_{0}\frac{\sin(\beta_{0,2}z)}{\sin(\beta_{0,2}h)}e^{j\beta_{0,1}h}e^{jk_{t_{0,1}}y}
\end{aligned}
\end{equation}
between the grid and the PEC ($-h<z<0$), where the longitudinal ($\beta_{0,i}$) and transverse ($k_{t_{0,i}}$) wavenumbers in the the two media ($i=1,2$) are given by
\begin{equation}
\label{eq:constants_E_ext}
\begin{aligned}
\beta_{0,1}=&k_{1}\cos \theta_{0,1}, &k_{t_{0,1}}&=k_{1}\sin \theta_{0,1}\\
\beta_{0,2}=&k_{2}\cos \theta_{0,2}, &k_{t_{0,2}}&=k_{2}\sin \theta_{0,2},
\end{aligned}
\end{equation}
where \textcolor{black}{the angles of propagation in the two media $\theta_{0,1}=\theta_\mathrm{in}$ and $\theta_{0,2}$ are related via Snell's law,
\begin{equation}
\label{eq:snell_law}
\begin{aligned}
n_{1}\sin \theta_{0,1}=n_{2}\sin \theta_{0,2},
\end{aligned}
\end{equation}
with $n_{1}=\sqrt{\varepsilon_1}$ and $n_{2}=\sqrt{\varepsilon_2}$ being the refractive indices of medium 1 and 2, respectively.} 
The reflection and transmission coefficients in \eqref{eq:E_ext_air} and \eqref{eq:E_ext_dielectric} are given\textcolor{black}{, respectively,} by
\begin{equation}
\label{eq:Reflection and transmission coefeicients}
\begin{aligned}
R_{0}=&\frac{j\gamma_{0}\tan(\beta_{0,2}h)-1}{j\gamma_{0}\tan(\beta_{0,2}h)+1}\\ T_{0}=&1+R_{0}=\frac{2j\gamma_{0}\tan(\beta_{0,2}h)}{j\gamma_{0}\tan(\beta_{0,2}h)+1},
\end{aligned}
\end{equation}
with the wave-impedance ratio $\gamma_0$ defined as
\begin{equation}
\label{equ:impedance_ratio}
\begin{aligned}
\gamma_{0}=\frac{\eta_{2}}{\eta_{1}}\frac{\cos \theta_{0,1}}{\cos \theta_{0,2}}.
\end{aligned}
\end{equation}
The subscripts in the above notations will become clear shortly, when we introduce the \textcolor{black}{FB} mode expansion of the fields, resulting from the grid periodicity.

Next, we derive the fields produced by the grid itself. We treat the current-carrying strips as secondary sources, and solve the resulting scattering problem by employing, once again, the suitable boundary conditions.
To formulate the grid-generated fields, we first introduce an infinitesimal layer in medium 1 with a thickness of $\delta<<h,\lambda,$ and consider the grid to be positioned at $z=-h-\delta$ [Fig. \ref{metagrating}(b)]; to reproduce the actual configuration of Fig. \ref{metagrating}(a) we will eventually take $\delta\rightarrow0$ to eliminate this auxiliary layer. The field created by such a grid \textcolor{black}{in free space} can be approximated (for thin wires) by \textcolor{black}{the} field formed by an infinite series of electric line sources \cite{tretyakov2003analytical},
\begin{equation}
\label{eq:field_from_wire_source}
\begin{aligned}
&E^\mathrm{ls}_{x,1}(y,z)=\\&-\!\!\frac{k_{1}\eta_{1}}{4}I\!\!\!\!\sum_{n=-\infty}^{\infty}\!\!\!\!e^{-jk_{1}n\Lambda\sin \theta_\mathrm{in}}
H^{(2)}_{0}[k_{1}\!\sqrt{\!(\!z+h+\delta\!)^{2}\!+\!(\!y-n\Lambda\!)^{2}}]
\end{aligned}
\end{equation}
\textcolor{black}{where $I$ is the current induced in the wires due to the applied fields, yet to be evaluated.} 
As in \cite{tretyakov2003analytical}, we use the Poisson formula to express these line-source fields in medium 1 as a superposition of FB modes
\begin{equation}
\label{eq:Poisson_formula}
\begin{aligned}
&E^\mathrm{ls}_{x,1}(y,z)=
-\frac{k_{1}\eta_{1}}{2\Lambda}I\sum_{m=-\infty}^{\infty}\frac{e^{jk_{t_{m,1}}y}e^{-j\beta_{m,1}\mid z+h+\delta\mid}}{\beta_{m,1}},
\end{aligned}
\end{equation}
where the longitudinal ($\beta_{m,i}$) and transverse ($k_{t_{m,i}}$) wavenumbers are given by a generalization of \eqref{eq:constants_E_ext}
\begin{equation}
\label{eq:FB_wave_numbers}
\begin{aligned}
&k_{t_{m,i}}=k_{1}\sin \theta_\mathrm{in}+\frac{2\pi m}{\Lambda}=k_{i}\sin \theta_{m,i}\\
&\beta_{m,i}=\sqrt{k_{i}^{2}-\big(k_{1}\sin \theta_\mathrm{in}+\frac{2\pi m}{\Lambda}\big)^{2}}=k_{i}\cos \theta_{m,i}
\end{aligned}
\end{equation}
In \eqref{eq:FB_wave_numbers}, the subscript $m$ corresponds to the order of the FB mode under consideration, and the subscript $i$, as before, refers to the medium where the fields are evaluated ($i=1,2$).


Consequently, the fields produced by the grid in the presence of the substrate and the PEC can be written in general as
\begin{equation}
\label{eq:E_wire1_general}
\begin{aligned}
&E^\mathrm{grid}_{x,1}(y,z)=-\frac{k_{1}\eta_{1}}{2\Lambda}I\!\!\!\sum_{m=-\infty}^{\infty}\!\!A_{m,1}\frac{e^{jk_{t_{m,1}}y}e^{-j\beta_{m,1}(z+h+\delta)}}{\beta_{m,1}}\\
&-\frac{k_{1}\eta_{1}}{2\Lambda}I\!\!\!\sum_{m=-\infty}^{\infty}\!\!B_{m,1}\frac{e^{jk_{t_{m,1}}y}e^{+j\beta_{m,1}(z+h+\delta)}}{\beta_{m,1}}
\end{aligned}
\end{equation}
for $(-h-\delta)<z<-h$ (in medium 1), where we consider the scattering of the individual modes off the structure \cite{felsen1994radiation} (note that in the infinitesimal layer $\left|z+h+\delta\right|=z+h+\delta$). A similar expression is applicable for the fields inside the substrate ($-h<z<0$), 
\begin{equation}
\label{eq:E_wire2_general}
\begin{aligned}
&E^\mathrm{grid}_{x,2}(y,z)=-\frac{k_{2}\eta_{2}}{2\Lambda}I\!\!\!\sum_{m=-\infty}^{\infty}\!\!A_{m,2}\frac{e^{jk_{t_{m,2}}y}e^{-j\beta_{m,2}z}}{\beta_{m,2}}\\
&-\frac{k_{2}\eta_{2}}{2\Lambda}I\!\!\!\sum_{m=-\infty}^{\infty}\!\!B_{m,2}\frac{e^{jk_{t_{m,2}}y}e^{+j\beta_{m,2}z}}{\beta_{m,2}},
\end{aligned}
\end{equation}
\textcolor{black}{with the wavenumbers $(\beta_{m,2},k_{t_{m,2}})$ modified as per Snell's law \cite{wait1957impedance},} in consistency with the definitions \eqref{eq:FB_wave_numbers}.

The spectral expansions \eqref{eq:E_wire1_general} and \eqref{eq:E_wire2_general} express the fields between the grid and the PEC (in both media) as a superposition of plane waves [Fig. \ref{metagrating}(b)]: the amplitudes $A_{m,1},B_{m,1}$ correspond, respectively, to the forward (transmitted) and backward (reflected) \textcolor{black}{propagating} waves in medium 1 (the infinitesimal layer), while the amplitudes $A_{m,2},B_{m,2}$ correspond to the forward and backward waves in medium 2 (the substrate). Due to the orthogonality of these FB modes, the boundary conditions can be enforced individually for each plane wave to resolve the relations between the coefficients $A_{m,i},B_{m,i}$\cite{felsen1994radiation,Chew1990}.
Specifically, for observation points inside the infinitesimal layer ($-h-\delta<z<-h$) we eventually get 
\begin{equation}
\label{eq:E_wire1_infinitesimal_layer}
\begin{aligned}
&E^\mathrm{grid}_{x,1}(y,z)=-\frac{k_{1}\eta_{1}}{2\Lambda}I\!\!\!\sum_{m=-\infty}^{\infty}\!\!\frac{e^{-j\beta_{m,1}(z+h+\delta)}}{\beta_{m,1}}e^{jk_{t_{m,1}}y}\\
&-\frac{k_{1}\eta_{1}}{2\Lambda}I\!\!\!\sum_{m=-\infty}^{\infty}\!\!R_{m}\frac{e^{j\beta_{m,1}(z+h+\delta)}}{\beta_{m,1}}e^{jk_{t_{m,1}}y}
\end{aligned}
\end{equation}
while for the fields in the dielectric medium ($-h<z<0$) we get
\begin{equation}
\label{eq:E_wire_dielectric}
\begin{aligned}
E^\mathrm{grid}_{x,2}(y,z)=\frac{k_{2}\eta_{2}}{2\Lambda}I\!\!\!\sum_{m=-\infty}^{\infty}&\frac{T_{m}}{\gamma_{m}\beta_{m,2}}\frac{\sin\beta_{m,2}z}{\sin\beta_{m,2}h}e^{jk_{t_{m,1}}y}
\end{aligned}
\end{equation}
where the reflection and transmission coefficients of the $m$th mode are given by \textcolor{black}{a} generalization of \eqref{eq:Reflection and transmission coefeicients}
\begin{equation}
\label{eq:constants_E1_wire}
\begin{aligned}
&R_{m}=\frac{j\gamma_{m}\tan \beta_{m,2}h -1}{j\gamma_{m}\tan \beta_{m,2}h +1} \\
&T_{m}=1+R_{m}=\frac{2j\gamma_{m}\tan \beta_{m,2}h}{j\gamma_{m}\tan \beta_{m,2}h +1},
\end{aligned}
\end{equation}
and the wave-impedance ratio \textcolor{black}{is} generalized after \eqref{equ:impedance_ratio} to be
\begin{equation}
\label{eq:constants_E1_wire_gamma}
\begin{aligned}
\gamma_{m}=\frac{\eta_{2}}{\eta_{1}}\frac{\cos \theta_{m,1}}{\cos \theta_{m,2}}.
\end{aligned}
\end{equation}

The fields in the region $z<-h-\delta$ (medium 1) can be constructed from the fields in the infinitesimal layer \eqref{eq:E_wire1_infinitesimal_layer} by considering the symmetric nature of the line-source fields \eqref{eq:Poisson_formula} [first term of \eqref{eq:E_wire1_infinitesimal_layer}] and the continuity of the reflected fields [second term of \eqref{eq:E_wire1_infinitesimal_layer}] across the grid position $z=-h-\delta$. 
Consequently, utilizing $T_m=1+R_m$ of \eqref{eq:constants_E1_wire}, the field at $z<-h-\delta$ reads
\begin{equation}
\label{eq:E_wire_air}
\begin{aligned}
E^\mathrm{grid}_{x,1}(y,z)=-\frac{k_{1}\eta_{1}}{2\Lambda}I\sum_{m=-\infty}^{\infty}T_{m}\frac{e^{+j\beta_{m,1}(z+h+\delta)}}{\beta_{m,1}}e^{jk_{t_{m,1}}y}.
\end{aligned}
\end{equation}

Finally, we recall that in the actual configuration [Fig. \ref{metagrating}(a)] the conducting strips are defined \emph{on} the dielectric substrate, i.e. are positioned at $z=-h$. Therefore, to obtain the accurate fields everywhere in space we should take $\delta\rightarrow0$ and eliminate the auxiliary infinitesimal layer that served us during the formulation. Substituting $\delta\rightarrow0$ into \eqref{eq:E_wire_air} yields the fields generated by the current-carrying strips in the presence of the substrate and PEC in medium 1 ($z<-h$)
\begin{equation}
\label{eq:E_wire_air_delta_0}
\begin{aligned}
E^\mathrm{grid}_{x,1}(y,z)=-\frac{k_{1}\eta_{1}}{2\Lambda}I\sum_{m=-\infty}^{\infty}T_{m}\frac{e^{+j\beta_{m,1}(z+h)}}{\beta_{m,1}}e^{jk_{t_{m,1}}y}
\end{aligned}
\end{equation}
while the field in medium 2 ($0<z<-h$) is given by \eqref{eq:E_wire_dielectric}.
Naturally, the total fields in each medium would be a superposition of the external and grid-produced fields $E^\mathrm{tot}_{x,i}=E^\mathrm{ext}_{x,i}+E^\mathrm{grid}_{x,i}$, evaluated by \eqref{eq:E_ext_air}, \eqref{eq:E_ext_dielectric}, \eqref{eq:E_wire_dielectric}, and \eqref{eq:E_wire_air_delta_0}.

\subsection{Propagating mode selection rules}
\label{subsec:Propagating mode selection rules}
As observed in \eqref{eq:FB_wave_numbers}, in consistency with the FB theorem, the scattered field in medium 1 consists of a superposition of modes departing at angles $\theta_{m,1}$, given by
\begin{equation}
\label{eq:momentum_conservation}
\begin{aligned}
\sin \theta_{m,1}=\sin \theta_\mathrm{in}+\frac{2\pi m}{k_{1}\Lambda} , m=0,\pm 1,\pm 2 ...
\end{aligned}
\end{equation}
As was demonstrated in \cite{ra2017meta, epstein2017unveiling}, in order to facilitate perfect coupling of the incident plane wave $\theta_\mathrm{in}$ to the desirable reflected one $\theta_\mathrm{out}$ using a single meta-atomic degree of freedom, we need to ensure that such coupling is allowed by the FB theorem, and that only these two modes, corresponding to $m=0$ ($\theta_\mathrm{in}$) and $m=-1$ ($\theta_\mathrm{out}$), can carry radiative power.

The $m$th mode would be propagating only if $\theta_{m}$ of \eqref{eq:momentum_conservation} is real, namely, if 
$\left|\sin\theta_\mathrm{in}+m\frac{\lambda_{1}}{\Lambda}\right|<1$, where $\lambda_1=2\pi/k_1$ is the effective wavelength in medium 1. Subsequently, to guarantee that the only propagating mode (besides the fundamental $m=0$) will be the $m=-1$ mode, the grid period must satisfy \cite{ra2017meta}
\begin{equation}
\label{eq:Lambda_range}
\frac{\lambda_{1}}{1+\sin\theta_\mathrm{in}}<\Lambda<
  \begin{cases}
    \frac{\lambda_{1}}{1-\sin \theta_\mathrm{in}}      & \quad 0<\theta_\mathrm{in}<\sin^{-1}(\frac{1}{3})\\
    \frac{2\lambda_{1}}{1+\sin \theta_\mathrm{in}}      & \quad \sin^{-1}(\frac{1}{3})<\theta_\mathrm{in}<\frac{\pi}{2}\\
  \end{cases}
\end{equation}
At the same time, the grid period should match the condition for coupling the incident wave with angle $\theta_\mathrm{in}$ to the $m=-1$ FB mode propagating towards the desirable $\theta_\mathrm{out}$ \emph{in medium 1}.
In other words, we require that $\theta_{-1,1}=\theta_\mathrm{out}$, which, using \eqref{eq:momentum_conservation}, implies that
\begin{equation}
\label{eq:Lambda_theta_out_minus1}
\begin{aligned}
\sin \theta_\mathrm{in}-\sin \theta_\mathrm{out}=\frac{\lambda_{1}}{\Lambda}
\end{aligned}
\end{equation}
Finally, we can translate the condition over $\Lambda$ to a condition on $\theta_\mathrm{out}$, namely,
\begin{equation}
\label{eq:theta_out_range}
-\frac{\pi}{2}<\theta_\mathrm{out}<
  \begin{cases}
    \sin^{-1}(2\sin\theta_\mathrm{in}-1)   & \quad 0<\theta_\mathrm{in}<\sin^{-1}(\frac{1}{3})\\
    \sin^{-1}(\frac{1}{2}\sin \theta_\mathrm{in}-\frac{1}{2})  & \quad \sin^{-1}(\frac{1}{3})<\theta_\mathrm{in}<\frac{\pi}{2}\\
  \end{cases}
\end{equation}

The modes scattered into the substrate, however, depart at angles $\theta_{m,2}$, following the modal Snell's law [\textit{cf.} \eqref{eq:momentum_conservation}]
\begin{equation}
\label{eq:momentum_conservation in dielectric}
\begin{aligned}
k_{2}\sin \theta_{m,2}=k_{1}\sin \theta_\mathrm{in}+\frac{2\pi m}{\Lambda} , m=0,\pm 1,\pm 2 ...
\end{aligned}
\end{equation}
Hence, for the $m$th FB mode to be propagating in \emph{the dielectric}, the mode index should satisfy
$\left|\frac{n_{1}}{n_{2}}\sin\theta_\mathrm{in}+\frac{\lambda_{2}m}{\Lambda}\right|<1$ where $\lambda_2=2\pi/k_2$ is the effective wavelength in medium 2, namely,
\begin{equation}
\label{eq:condition for modes in dielectric}
\begin{aligned}
-\frac{\Lambda}{\lambda_{0}}(n_{2}+n_{1}\sin \theta_\mathrm{in})<m<\frac{\Lambda}{\lambda_{0}}(n_{2}-n_{1}\sin \theta_\mathrm{in})
\end{aligned}
\end{equation}
In practice, this result implies that determining the period $\Lambda$ according to \eqref{eq:Lambda_theta_out_minus1} to guarantee that only the $m=0$ and $m=-1$ modes are propagating in medium 1 does not necessarily prevent higher-order modes to be propagating in medium 2. The number of such propagating modes trapped in the substrate can be evaluated via \eqref{eq:condition for modes in dielectric}. 

\subsection{Eliminating specular reflection}
\label{subsec:Eliminating specular reflection}
Choosing $\theta_\mathrm{in}$, $\theta_\mathrm{out}$, and $\Lambda$ following \eqref{eq:Lambda_range} and \eqref{eq:Lambda_theta_out_minus1} guarantees that only the $m=0$ (specular) and $m=-1$ (anomalous) reflected FB modes can carry radiated power away from the metagrating. Hence, to ensure exclusive coupling of the incident power to the desirable anomalous reflection mode, we merely need to eliminate the power coupled to specular reflection. This is achieved by tuning the metagrating configuration such that the specular component of the external field is cancelled by the respective (zeroth-order) modal component of the field generated by the grid at $z\rightarrow-\infty$.
These are given, respectively, by the second term in \eqref{eq:E_ext_air} and the $m=0$ term in the summation \eqref{eq:E_wire_air_delta_0}. Therefore, the specular reflection elimination condition can be formulated by demanding the sum of these two terms to vanish, namely,
\begin{equation}
\label{eq:mode0_cancellation}
\begin{aligned}
E_\mathrm{in}R_{0}&e^{+j\beta_{0,1}(z+2h)}e^{jk_{t_{0,1}}y}\\
&-\frac{k_{1}\eta_{1}}{2\Lambda}IT_{0}\frac{e^{+j\beta_{0,1}(z+h)}}{\beta_{0,1}}e^{jk_{t_{0,1}}y}=0
\end{aligned}
\end{equation}
In other words, in order to ensure that no power is coupled to the reflected $m=0$ mode, we should set the design degrees of freedom ($\tilde{Z}$ and $h$) such that the current induced on the strips would follow
\begin{equation}
\label{eq:Condition_specular_cancellation}
\begin{aligned}
I=\frac{2E_\mathrm{in}\Lambda \cos \theta_\mathrm{in}e^{jk_{1}h\cos \theta_\mathrm{in}}}{\eta_{1}}\frac{R_{0}}{1+R_{0}},
\end{aligned}
\end{equation}
where $R_{0}$ is given by \eqref{eq:Reflection and transmission coefeicients}.

\subsection{Perfect anomalous reflection}
\label{subsec:power_conservation}
Satisfaction of the specular reflection elimination condition \eqref{eq:Condition_specular_cancellation} guarantees that the only propagating reflected mode is the $m=-1$ FB mode,
directed towards $\theta_\mathrm{out}$. 
Hence, to achieve an optimal (unitary) coupling efficiency between the incident fields to this mode, what is left is to ensure that no power is absorbed by the grating. This requirement is equivalent to demanding that the net real power \textcolor{black}{crossing} a given plane ${z<-h}$ would vanish \cite{epstein2017unveiling}. Such a result would imply that all the incident power is reflected from the metagrating, indicating that the structure is passive and lossless.
Formally, this power conservation condition reads
\begin{equation}
\label{eq:real_power}
\begin{aligned}
P_{z}(z)=\frac{1}{2}\int\limits_{-\Lambda/2}^{\Lambda/2} \Re\{E_{x,1}(y,z)H^{*}_{y,1}(y,z)\}dy=0
\end{aligned}
\end{equation}
where we \textcolor{black}{relied on} the configuration periodicity to restrict the integration limits to a single period.

Using equations \eqref{eq:E_ext_air} and \eqref{eq:E_wire_air_delta_0}, the total field at the plane $z<-h$ can be written as
\begin{equation}
\label{eq:total_field_mode_minus1}
\begin{aligned}
&E_{x,1}^\mathrm{tot}(y,z<-h)=E_{0}e^{-j\beta_{0,1}z}e^{jk_{t_{0,1}}y}\\
&-\frac{k_{1}\eta_{1}}{2\Lambda}I\cdot T_{-1}\frac{e^{+j\beta_{-1,1}(z+h)}}{\beta_{-1,1}}e^{jk_{t_{-1,1}}y}\\
&-j\frac{k_{1}\eta_{1}}{2\Lambda}I\!\!\!\sum_{m=-\infty}^{\infty}\!\!T_{m}\frac{e^{\alpha_{m,1}(z+h)}}{\alpha_{m,1}}e^{jk_{t_{m,1}}y}
\end{aligned}
\end{equation}
where we used the notation $\beta_{m,1}\triangleq-j\alpha_{m,1}\,\,(\alpha_{m,1}\geq0)$ for the evanescent higher-order modes $m\neq0,-1$.
Inserting the total electric field (\ref{eq:total_field_mode_minus1}) and the corresponding magnetic field $\left[H_{y}(y,z)=-\frac{1}{jk\eta_1}\frac{\partial}{\partial z} E_{x}(y,z)\right]$ into \eqref{eq:real_power}, and incorporating the specular reflection elimination condition \eqref{eq:Condition_specular_cancellation}, yields the condition for achieving perfect anomalous reflection, namely
\begin{equation}
\label{eq:final_condition}
\begin{aligned}
\frac{\cos \theta_\mathrm{out}}{\cos \theta_\mathrm{in}}=\frac{\left|1+R_{-1}\right|^{2}}{\left|1+R_{0}\right|^{2}},
\end{aligned}
\end{equation}
where $R_0$ and $R_{-1}$ are given by \eqref{eq:constants_E1_wire}.
\textcolor{black}{Examination of \eqref{eq:constants_E1_wire} reveals that} for given $\theta_\mathrm{in}$, $\theta_\mathrm{out}$, $\varepsilon_1$, and $\varepsilon_2$, \textcolor{black}{the condition \eqref{eq:final_condition}} forms a nonlinear equation for the substrate thickness $h$, typically resolved \textcolor{black}{by a simple numerical code} \cite{ra2017meta,epstein2017unveiling}. 

It is interesting to note that in the special case where $\varepsilon_{1}=\varepsilon_{2}$ (homogeneous medium), \textcolor{black}{the condition} \eqref{eq:final_condition} reduces to
\begin{equation}
\label{eq:final_condition_special_case}
\begin{aligned}
\frac{\cos \theta_\mathrm{out}}{\cos \theta_\mathrm{in}}=\frac{\sin^{2}(k_{1}h\cos \theta_\mathrm{out})}{\sin^{2}(k_{1}h\cos \theta_\mathrm{in})},
\end{aligned}
\end{equation}
which is consistent with the analogous condition derived in \cite{ra2017meta} for obtaining perfect anomalous reflection for transverse magnetic (TM) polarized plane waves with a magnetically-polarizable metagrating embedded in homogeneous medium. The only difference is that in \cite{ra2017meta}, image theory for TM-polarized sources dictates the trigonometric functions on the right hand side to be cosines, whereas for our TE-polarization case, sine functions describe the relevant interference phenomena.

\subsection{Distributed load impedance}
\label{subsec:Distributed load impedance}
Once the distance $h$ between the wire array and the PEC has been determined via (\ref{eq:final_condition}), \textcolor{black}{fixing} our first degree of freedom, we proceed to \textcolor{black}{setting} the second degree of freedom in our device, namely, the load impedance.
To this end, we formulate the total field on a reference wire at the position $(y,z)\rightarrow(0,-h)$ and utilize Ohm's law, $E_{x,1}^\mathrm{tot}(y\rightarrow0,z\rightarrow-h)=\widetilde Z I$, to assess the required impedance per unit length $\widetilde Z$ \cite{epstein2017unveiling,tretyakov2003analytical}.

This total field is composed of the fields given by equations (\ref{eq:E_ext_air}) and \eqref{eq:E_wire_air_delta_0}. While the external field at the reference strip is given by substituting $(y,z)=(0,-h)$ in (\ref{eq:E_ext_air}), \textcolor{black}{this substitution cannot be simply used in the summation \eqref{eq:E_wire_air_delta_0} to evaluate the field generated by the grid at that position; this is because} the Hankel function diverges at this point, rendering the Poisson formula \textcolor{black}{used to arrive at (\ref{eq:E_wire_air_delta_0})} invalid. 
\textcolor{black}{Instead, we must isolate the singular terms in \eqref{eq:E_wire_air_delta_0}, related to the self-fields induced on the reference strip by the current flowing through it, retrace our steps and use a refined modal representation that allows proper evaluation of these terms (see Appendix \ref{sec:App}). Combining the two field contributions, Ohm's law reads \cite{epstein2017unveiling,tretyakov2003analytical}}
\begin{equation}
\label{eq:equation_on_load}
\begin{aligned}
&\widetilde{Z} I = E_\mathrm{in}(e^{jk_{1}h\cos \theta_\mathrm{in}}+R_{0}e^{jk_{1}h\cos \theta_\mathrm{in}})\\
&-\frac{k_{1}\eta_{1}}{2\Lambda}I\sum_{m=-\infty}^{\infty}\frac{R_m}{\beta_{m,1}}-\frac{k_{1}\eta_{1}I}{2}\Big\{\frac{1}{k_{1}\Lambda \cos \theta_\mathrm{in}}+\frac{j}{\pi}\Big[\log \frac{\Lambda}{2\pi r_\mathrm{eff}}\\
&+\frac{1}{2}\sum_{\substack{m=-\infty \\ m\neq 0}}^{\infty}\Big(\frac{2\pi}{\sqrt{(2\pi m+k_{1}\sin \theta_\mathrm{in})^{2}-(k\Lambda)^{2}}}-\frac{1}{\left| m \right|}\Big)\Big]\Big\}
\end{aligned}
\end{equation}
and we recall that the effective radius is $r_\mathrm{eff}=w/4$ \cite{tretyakov2003analytical}.

Finally, after substituting equations \eqref{eq:Condition_specular_cancellation} and \eqref{eq:final_condition} into \eqref{eq:equation_on_load}, we \textcolor{black}{find} the load impedance per unit length required to fully couple the incoming plane wave into the desirable FB mode \textcolor{black}{to be}
\begin{equation}
\label{eq:final_load_impedance}
\begin{aligned}
\widetilde Z=&-j\frac{\eta_{1}\left|1+R_{-1}\right|^{2}}{4\Lambda \cos \theta_\mathrm{out}}\left[\frac{1}{\gamma_{0}\tan \beta_{0,2}h}+\frac{1}{\gamma_{-1}\tan \beta_{-1,2}h}\right]\\
&+j\frac{k_{1}\eta_{1}}{2\pi}\left(\frac{1}{2}+\log \frac{2\pi r_\mathrm{eff}}{\Lambda}\right)\\
&-j\frac{\eta_{1}}{\Lambda}\sum_{\substack{m=-\infty \\ m\neq 0,-1}}^{\infty}\left[\frac{k_{1}(1+R_{m})}{2\alpha_{m,1}}-\frac{k_{1}\Lambda}{4\pi}\frac{1}{\left| m \right|}\right].
\end{aligned}
\end{equation}
It is apparent that the resultant impedance load is purely imaginary, which is expected given that we designed our structure to be passive and lossless by enforcing power conservation in Section \ref{subsec:power_conservation}.
Consequently, for given incidence and desirable reflection angles, and the substrate properties, using a substrate thickness that satisfies \eqref{eq:final_condition} and setting the load impedance following \eqref{eq:final_load_impedance} would yield a passive dielectric-supported metagrating that deflects \emph{all} the power incoming from $\theta_\mathrm{in}$ towards $\theta_\mathrm{out}$.

\section{Results and discussion}
\label{sec:results}
To demonstrate and verify our synthesis methodology, we follow the prescribed procedure to design a PCB metagrating defined on a prototypical PEC-backed low-loss microwave substrate, having dielectric permittivity $\varepsilon_{2}=\varepsilon_{sub}=3\varepsilon_0$ ($\varepsilon_0$ is the vacuum permittivity) and loss tangent $\tan \delta=0.001$. For the conductors forming the wire grid we consider a strip width of $w=3\mathrm{mil}=76.2\mathrm{\mu m}$ and metal thickness of $t=18\mathrm{\mu m}$ [Fig. \ref{metagrating}(c)], compatible with standard fabrication capabilities\cite{epstein2016cavity}. The operating frequency is chosen to be $f=10\mathrm{GHz}$; to form a homogenous current distribution along the $x$ axis, we choose the repetition period of the lumped loads to be $L=\lambda/10\ll\lambda$ along this dimension.

The metagrating is designed to implement anomalous reflection of a plane wave incident at $\theta_\mathrm{in}=10^{\circ}$ towards $\theta_\mathrm{out}$, where this designated output angle is varied in the range $-85^{\circ}$ to $-45^{\circ}$, satisfying \eqref{eq:Lambda_theta_out_minus1}. 
Our first step is to find $h$, the required thickness of the dielectric substrate, for any of the desired output angles $\theta_\mathrm{out}$.
To this end, we define $\rho$ to quantify the deviation from the perfect anomalous reflection condition (\ref{eq:final_condition}), namely,
\begin{equation}
\label{eq:rho_equation}
\begin{aligned}
\rho\triangleq\frac{\cos \theta_\mathrm{out}}{\cos \theta_\mathrm{in}}-\frac{\left|1+R_{-1}\right|^{2}}{\left|1+R_{0}\right|^{2}},
\end{aligned}
\end{equation}
and we seek the $h$ values that minimizes $\left|\rho\right|$, for a given value of $\theta_\mathrm{out}$.

\begin{figure}[t]
\centering
\includegraphics[width=3.0in]{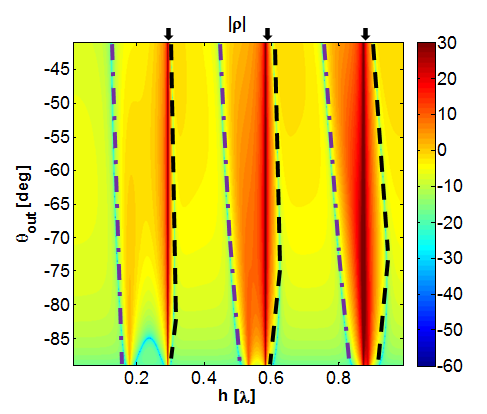}
\caption{\textcolor{black}{Deviation from the perfect anomalous reflection condition $\left|\rho\right|$, defined in \eqref{eq:rho_equation} (in dB scale)}, as a function of \textcolor{black}{the substrate thickness} $h$ and \textcolor{black}{the reflection angle} $\theta_\mathrm{out}$, for the \textcolor{black}{metagrating configuration specified} in Section \ref{sec:results}. \textcolor{black}{The dash-dotted and dashed lines indicate the solution branches of \eqref{eq:final_condition}, where the deviation $\left|\rho\right|$ is minimal. The black dashed lines represent working points that are close to the destructive interference points (marked with black arrows on the top of the plot), and thus are more prone to losses; the purple dash-dotted lines, on the other hand, are further away from these problematic thicknesses, forming the preferable solution branches (see discussion in Section \ref{sec:results}).}}
\label{Fig:Theta_out_h}
\end{figure}

Fig. \ref{Fig:Theta_out_h} presents a 2D plot of the deviation $\left|\rho\right|$ as a function of $h$ and $\theta_\mathrm{out}$, in logarithmic scale. Due to the nonlinear nature of \eqref{eq:final_condition}, several solution branches exist (marked in \textcolor{black}{dash-dotted purple and dashed black} lines).  
Although it seems that any branch would suffice for satisfying the condition and setting the design parameters, the different branches lead to completely different performance when it comes to realistic devices. As was discussed in detail in \cite{epstein2017unveiling}, when the metagrating configuration is chosen such that the incident and reflected fields destructively interfere on the grid position $z=h$, very high currents will be induced on the conductors (by design) in order to eliminate specular reflection. Under these operating conditions, even minor conductor losses, which are inevitable, may lead to significant power dissipation, and consequently considerable performance reduction; the same working points also exhibit reduced fractional bandwidth \cite{epstein2017unveiling}. Examining \eqref{eq:Condition_specular_cancellation}, it is observable that this high-current situation occurs for these working points where the denominator $(1+R_0)$ tends to 0, i.e., when $\tan(\beta_{0,2}h)\rightarrow0$ [see \eqref{eq:Reflection and transmission coefeicients}]. 

The discussion in the previous paragraph implies that we should avoid choosing solution branches for which the substrate thickness $h$ approaches the destructive interference values $h_\mathrm{destruct}$
\begin{equation}
\label{eq:h_loss}
\begin{aligned}
h_\mathrm{destruct}=\frac{\lambda}{2n_{2}\cos \theta_{0,2}}p, \,\,\,\,\,\,\,\,\, p=0,\pm 1,\pm 2 ...
\end{aligned}
\end{equation}
These thicknesses appear in Fig. \ref{Fig:Theta_out_h} as vertical high-value (red) lines at $h=0.29\lambda$, $h=0.58\lambda$, and $h=0.87\lambda$, marked with black arrows.
Therefore, any solution branch that is rather close to one of these vertical lines does not represent a good working point; these less preferable branches are marked in Fig. \ref{Fig:Theta_out_h} with dashed black lines. In \textcolor{black}{principle}, we \textcolor{black}{may} select any of the other solution branches (marked with \textcolor{black}{dash-dotted purple} lines) to set our metagrating configuration. They are sufficiently away from the destructive interference lines, thus leading to a moderate current in the strips that should not result in large losses.
Nonetheless, in order to achieve the most compact design, we choose herein the first "allowed" (\textcolor{black}{dash-dotted purple}) branch, featuring the smallest substrate thicknesses. 

Subsequently, we substitute the distances $h$ extracted from Fig. \ref{Fig:Theta_out_h} for each of the reflection angles $\theta_\mathrm{out}$ under consideration into (\ref{eq:final_load_impedance}) and evaluate the required \textcolor{black}{(purely-reactive)} distributed load impedance $\widetilde{Z}$. These impedances are found to be capacitive $\Im\{\widetilde Z\}<0$, corresponding to a lumped capacitor whose capacitance is given by the standard relation \cite{epstein2017unveiling}
\begin{equation}
\label{eq:capacitor formula}
\begin{aligned}
C=-1/(2\pi fL\Im \{\widetilde Z\}).
\end{aligned}
\end{equation}

\begin{table*}[t]
\centering
\begin{threeparttable}[b]
\renewcommand{\arraystretch}{1.3}
\caption{Design specifications and simulated performance of the designed metagratings operating at $f=10\mathrm{GHz}$ (corresponding to Fig. \ref{Fig:Width_theta_out} and \ref{Fig:Field plot}).}
\label{tab:metagrating_performance_10GHz}
\centering
\begin{tabular}{l|c|c|c|c|c|c|c|c|c|l}
\hline \hline
$\theta_\mathrm{out}$ 
& $-85^\circ$ & $-80^\circ$ & $-75^\circ$ 
& $-70^\circ$ & $-65^\circ$ & $-60^\circ$ 
& $-55^\circ$ & $-50^\circ$ & $-45^\circ$\\ 
\hline \hline \\[-1.3em]
	\begin{tabular}{l} $\Lambda [\lambda]$ \end{tabular}
	 & $0.854$ & $0.862$ & $0.876$ 
	& $0.897$ & $0.925$ & $0.961$ 
	& $1.006$ & $1.063$ &1.134 \\	\hline	
	 \begin{tabular}{l} $h [\lambda]$ \end{tabular}
	 & $0.153$ & $0.146$ & $0.14$ 
	& $0.136$ & $0.133$ & $0.13$ 
	& $0.127$ & $0.125$  &0.123 \\	\hline
	 \begin{tabular}{l} $C [\mathrm{fF}]$ \end{tabular}
	 & $47.65$ & $54.01$ & $57.78$ 
	& $60.13$ & $61.78$ & $62.94$ 
	& $63.66$ & $63.95$ &63.62\\	\hline 	   
	 \begin{tabular}{l} $W [\mathrm{mm}]$ \end{tabular}
	 & $1.431$ & $1.622$ & $1.736$ 
	& $1.806$ & $1.856$ & $1.891$ 
	& $1.912$ & $1.93$ &1.911 \\	\hline 	  
	 \begin{tabular}{l} Anomalous reflection \end{tabular}
	 & $95.8\%$ & $95.6\%$ & $96.6\%$ 
	& $97.4\%$ & $97.2\%$ & $97.16\%$ 
	& $96.8\%$ & $95.9\%$ &95.3\% \\	\hline 
	\begin{tabular}{l} Specular reflection \end{tabular}
	 & $0.5\%$ & $1.69\%$ & $0.93\%$ 
	& $0.15\%$ & $0.71\%$ & $0.1\%$ 
	& $0.67\%$ & $1.38\%$ &$1.73\%$ \\	\hline 
	 \begin{tabular}{l} Losses \end{tabular}
	 & $3.7\%$ & $2.71\%$ & $2.47\%$ 
	& $2.45\%$ & $2.09\%$ & $2.74\%$ 
	& $2.53\%$ & $2.72\%$  &2.97\%\\	\hline
	\begin{tabular}{l} Bandwidth \end{tabular}
	 & $0.8\%$ & $4.39\%$ & $13.37\%$ 
	& $20.66\%$ & $17.05\%$ & $15.24\%$ 
	& $12.23\%$ & $9.53\%$  &6.8\%\\		
\hline \hline
\end{tabular}
\end{threeparttable}
\end{table*}

However, as we aim at devising a full PCB-compatible design, these lumped capacitances have to be implemented eventually as metallic traces. To this end, we harness the physical structure utilized in \cite{epstein2017unveiling} to realize a printed capacitor [Fig. \ref{metagrating}(c)], with the separation between the capacitor "plates" given by $s=w=3\mathrm{mil}=76.2\mathrm{\mu m}$, and the capacitor width denoted by $W$. In \cite{epstein2017unveiling} it was shown (following \cite{guptamicrostrip}) that for printed capacitors situated in free space (vacuum), the capacitor width corresponding to the required capacitance value $C$ can be evalutated using $W\approx2.85K_\mathrm{corr}C[\frac{\mathrm{mil}}{\mathrm{fF}}]$, where $K_\mathrm{corr}$ is a frequency-dependent correction value, found therein to be $K_\mathrm{corr}=0.83$ at $f=10\mathrm{GHz}$.
Nevertheless, in our configuration [Fig. \ref{metagrating}(a)], which considers the capacitor to be printed on a dielectric substrate, this formula has to be generalized to account for the permittivity of both media $1$ and $2$. 
We follow \cite{yakovlev2009analytical}, and approximate the effective permittivity near the interface as $\varepsilon_\mathrm{eff}=\frac{\varepsilon_{1}+\varepsilon_{2}}{2}$. Subsequently, the capacitance formula can be adjusted to consider the effective dielectric between the capacitor "plates", yielding
\begin{equation}
\label{eq:width_formula}
\begin{aligned}
W\approx\frac{2.85K_\mathrm{corr}C}{\varepsilon_\mathrm{eff}}[\frac{\mathrm{mil}}{\mathrm{fF}}].
\end{aligned}
\end{equation}
This step finalizes our design procedure, deriving all the required physical parameters of the metagrating for any desirable $\theta_\mathrm{out}$ in a semianalytical manner, without requiring even a single simulation in a commercial solver.


We verify these designs via full-wave simulations, conducted with CST Microwave Studio. We use a unit cell periodic boundary conditions in the frequency domain solver to simulate a single period of the metagrating, as depicted in Fig. \ref{metagrating}(c). 
The periodicity in the $x$ and $y$ directions is $L$ and $\Lambda$, respectively, and we use copper traces with a realistic conductivity of $\sigma=58\times10^{6}\mathrm{S/m}$ for both the printed capacitors and the strips.


\textcolor{black}{As presented earlier in this section,} the minimal substrate thicknesses $h$ that satisfy \eqref{eq:final_condition} are extracted from Fig. \ref{Fig:Theta_out_h} for several $\theta_\mathrm{out}$ values in the range $-45^\circ$ to $-85^\circ$, separated by $5^\circ$ from each other. Substituting these values into \eqref{eq:final_load_impedance}, we evaluate the required distributed load impedance $\widetilde{Z}$ for each desirable $\theta_\mathrm{out}$, and utilize \eqref{eq:capacitor formula} to assess the corresponding lumped capacitance $C$. Finally, we follow \eqref{eq:width_formula} to retrieve the width $W$ of the printed capacitor that would implement the required capacitance in practice, and then define the physical structure in CST Microwave Studio in order to probe the metagrating's response when illuminated by the \textcolor{black}{designated plane-wave excitation,} incoming from $\theta_\mathrm{in}=10^\circ$.

The results of these full-wave simulations are summarized in \textcolor{black}{Table} \ref{tab:metagrating_performance_10GHz}, along with the metagrating parameters as evaluated following the analytical design procedure. The anomalous reflection efficiency denoted in the table is defined as the ratio between the power reflected towards $\theta_\mathrm{out}$ (the $m=-1$ FB mode) and the incident power; the specular reflection efficiency is defined as the ratio between the power reflected towards $\theta_\mathrm{in}$ (the reflected $m=0$ FB mode) and the incident power; the losses are calculated as the difference between the sum of these two \textcolor{black}{parameters} and unity. As can be observed, the designed realistic PCB metagratings achieve very high anomalous reflection efficiencies ($>95\%$) even for wide-angle deflections, limited only by inevitable \textcolor{black}{realistic} conductor losses, which are fully taken into account in the simulation. As predicted by the theoretical model, specular reflection is practically eliminated and near-unity coupling efficiencies are reached without any optimization whatsoever.

In consistency with the discussion in the beginning of this section, choosing a solution branch that is sufficiently away from the destructive interference working points $h_\mathrm{destruct}$ indeed leads to very minor absorption losses ($<5\%$), for all considered $\theta_\mathrm{out}$. This result confirms the hypothesis presented in \cite{epstein2017unveiling}, where it was suggested that utilizing a dielectric substrate would allow low-loss diffraction engineering via metagratings for a wide range of angles, in contrast to metagratings positioned in homogeneous media, for which certain angles $\theta_\mathrm{out}$ have shown to be particularly prone to high losses. This is due to the spatial dispersion associated with the reflection from the PEC-backed substrate \eqref{eq:Reflection and transmission coefeicients}, which is capable of shifting the points of destructive interference away from these problematic $\theta_\mathrm{out}$ values.

The last row in Table \ref{tab:metagrating_performance_10GHz} indicates the fractional bandwidth exhibited by each of the considered metagratings. Similar to previous reports, \cite{epstein2017unveiling,wong2017perfect}, we define the fractioal bandwidth (BW) as the ratio between the frequency interval $\Delta f$ in which at least 90\% ($-0.46\mathrm{dB}$) of the power is coupled to the anomalous reflection mode, and the nominal operating frequency ($f=10\mathrm{GHz}$). The recorded values show that a moderate BW can be achieved for a wide range of reflection angles, in consistency with the observations in \cite{epstein2017unveiling,wong2017perfect}. 
Nonetheless, when $\theta_\mathrm{out}$ approaches the edges of the angular range defined following \eqref{eq:theta_out_range}, i.e. when $\theta_\mathrm{out}\rightarrow-85^\circ$ and $\theta_\mathrm{out}\rightarrow-45^\circ$, the BW deteriorates due to the fact that minor changes in the illumination frequency drives the reflection angle outside the allowed range, deviating from the formulation framework in which only the $m=0,-1$ FB modes are propagating. In these cases, when the frequency is slightly modified, either the $m\!=\!-1$ mode becomes evanescent ($\theta_\mathrm{out}\!\!\rightarrow\!-85^\circ$) or additional FB propagating modes appear ($\theta_\mathrm{out}\!\!\rightarrow\!\!-45^\circ$), reducing the coupling to the desirable anomalous reflection mode.

\begin{figure}[t]
\centering
\includegraphics[width=2.9in]{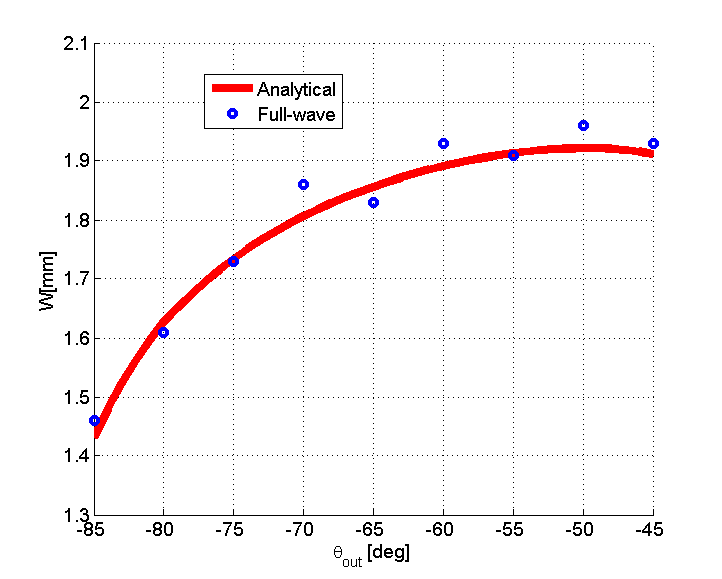}
\caption{Predicted values of the capacitor's width \textcolor{black}{(red solid line),} calculated from \eqref{eq:width_formula} as function of $\theta_\mathrm{out}$ following the detailed design scheme \textcolor{black}{(Table \ref{tab:metagrating_performance_10GHz})}, compared with optimization results using full-wave simulations \textcolor{black}{(blue circles)}.}
\label{Fig:Width_theta_out}
\end{figure}

To further assess the accuracy of the analytical model, we sweep the width of the printed capacitor $W$ in simulation and record the values that yield the highest anomalous reflection efficiency (as was done in \cite{ra2017meta} to finalize the metagrating design). These values are presented in blue circles in Fig. \ref{Fig:Width_theta_out} as a function of the designated reflection angle $\theta_\mathrm{out}$; for comparison, the analytically predicted $W$ are presented on the same plot using a solid red line. The comparison clearly shows that the analytical formulas \eqref{eq:final_load_impedance}, \eqref{eq:capacitor formula}, and \eqref{eq:width_formula} succeed very well in deriving the optimal geometrical parameters of the meta-atoms, demonstrating the high fidelity of the presented synthesis scheme.

\begin{figure}[t]
\centering
\includegraphics[width=3.5in]{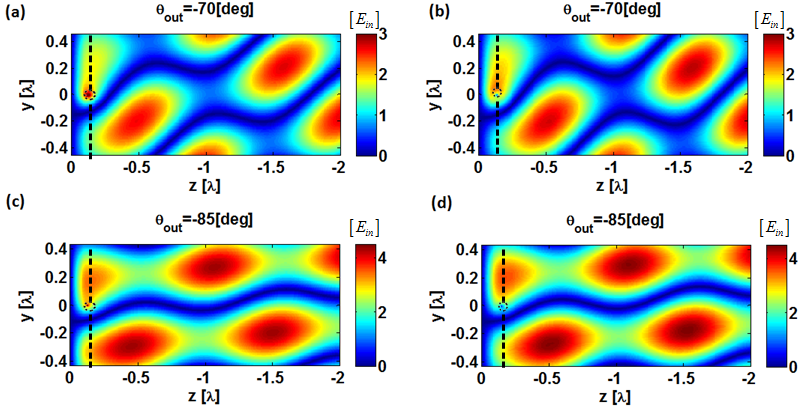}
\caption{Electric Field dtstributions $\mid \Re\{E_{x}(y,z)\}\mid$ for realistic metagrating comprised of loaded conducting strips on a PEC backed dielectric substrate. The dielectric constant of the substrate is $\varepsilon_\mathrm{sub}=3\varepsilon_0$ with loss tangent of $\tan \delta=0.001$. The incident angle is $\theta_\mathrm{in}=10^{\circ}$ and the metagrating operates at $f=10GHz.$ The analytical calculation following \textcolor{black}{\eqref{eq:E_wire_dielectric}, \eqref{eq:E_wire_air_delta_0}, \eqref{eq:Condition_specular_cancellation}, and \eqref{eq:final_condition}} [(a),(c)] is compared to full-wave simulations \textcolor{black}{with the configurations presented in Table \ref{tab:metagrating_performance_10GHz}} [(b),(d)]. A single period \eqref{eq:Lambda_theta_out_minus1} is presented for output angles of $\theta_\mathrm{out}=-70^{\circ}$ [(a),(b)] and $\theta_\mathrm{out}=-85^{\circ}$ [(c),(d)]. Dashed black vertical lines \textcolor{black}{denote the metagrating plane, i.e. $z=-h$, and the black dashed circles mark the region around the meta-atoms where deviation between the analytical prediction and the full-wave simulations is expected \cite{epstein2017unveiling}}.}
\label{Fig:Field plot}
\end{figure}

Finally, we compare the field distribution, analytically calculated using  \eqref{eq:E_wire_dielectric}, \eqref{eq:E_wire_air_delta_0}, \eqref{eq:Condition_specular_cancellation}, and \eqref{eq:final_condition}, to the fields as obtained from full-wave simulations for the metagratings defined using the optimal realistic printed capacitor extracted from Fig. \ref{Fig:Width_theta_out}. The field plots are presented in Fig. \ref{Fig:Field plot} for two representative reflection angles: $\theta_\mathrm{out}=-70^\circ$ [Fig. \ref{Fig:Field plot}(a),(b)] and $\theta_\mathrm{out}=-80^\circ$ [Fig. \ref{Fig:Field plot}(c),(d)]. The excellent correspondence between the analytical prediction and the full-wave simulation of the physical structure points out the impressive ability of the theoretical model to capture all the relevant wave phenomena in this scattering scenario. The only exception relates to the fields very close to the meta-atom position $\left(y,z\right)\approx\left(0,-h\right)$, marked in black dashed circles in Fig. \ref{Fig:Field plot}, where the analytical model, which considers the load impedance to be uniformly distributed along the current-carrying strips, fails to account accurately for the actual finite-size printed capacitor \cite{epstein2017unveiling}.

These results verify that the proposed device is capable of implementing highly-efficient anomalous reflection for a \textcolor{black}{broad} range of angles, properly described by an analytical model that allows full design of a PCB fabrication-ready metagrating structure, without requiring the usage of full-wave simulators.

\section{Conclusion}
To conclude, we have presented an analytical scheme for designing PCB metagratings for perfect anomalous reflection, comprising only a single meta-atom per period. The devices are composed of loaded conducting strips defined on a dielectric substrate backed by a PEC layer, thus allowing a realistic configuration suitable for production with standard (PCB) manufacturing techniques. The formulation enables evaluation of the appropriate substrate thickness and the suitable dimensions of the printed capacitor that facilitate reflection of a given incident plane wave towards a desirable non-specular direction. This is achieved using considerations of destructive interference phenomena to eliminate spurious specular reflections, and power conservation to guarantee unitary coupling efficiencies via a passive and lossless structure.

We demonstrated that by selecting proper working points, oriented by the observations reported in previous work \cite{epstein2017unveiling}, near-optimal wide-angle anomalous reflection can be achieved even in the presence of realistic losses. Devices operating at these preferable working points often feature relatively \textcolor{black}{large} bandwidth, in consistency with previous theoretical reports \cite{ra2017meta,epstein2017unveiling} and recent empirical evidence \cite{wong2017perfect}.
The analytical model was verified via full-wave simulations, confirming our design methodology for various design points. 
 
Importantly, our results reveal that the detailed rigorous analytical model can generate trustworthy fabrication-ready PCB designs without any full-wave optimization whatsoever, overcoming the considerable complexity involved with metasurface physical realization, which demands numerically-intensive simulations for the implementation of dense subwavelength meta-atom arrays. These \textcolor{black}{new analytical tools} would facilitate efficient synthesis of detailed metagrating designs, which would allow experimental characterization in future work, and are \textcolor{black}{expected} to accelerate the development of advanced metagratings for versatile diffraction engineering.

\appendices

\section{\textcolor{black}{Evaluating the grid-generated fields at the reference strip}}
\label{sec:App}

As was mentioned in section \ref{subsec:Distributed load impedance}, \textcolor{black}{it is not possible to} calculate the field generated from the grid at the reference strip by \textcolor{black}{simply} substituting $(y,z)=(0,-h)$ in \eqref{eq:E_wire_air_delta_0}\textcolor{black}{, as the summation} diverges at this point.
Instead, we \textcolor{black}{follow the formulation steps prescribed in this appendix to evaluate the fields at this position, necessary for relating the load impedance to the currents induced by the applied fields, and thus for finalizing the metagrating design.} 

\textcolor{black}{First, we isolate the singular components of \eqref{eq:E_wire_air_delta_0}, using \eqref{eq:constants_E1_wire} to divide it} into two terms as follows
\begin{equation}
\label{eq:E_wire_air_delta_0_Appendix}
\begin{aligned}
E^\mathrm{grid}_{x,1}(y,z)&=-\frac{k_{1}\eta_{1}}{2\Lambda}I\sum_{m=-\infty}^{\infty}\frac{e^{+j\beta_{m,1}(z+h)}}{\beta_{m,1}}e^{jk_{t_{m,1}}y}\\
&-\frac{k_{1}\eta_{1}}{2\Lambda}I\sum_{m=-\infty}^{\infty}R_{m}\frac{e^{+j\beta_{m,1}(z+h)}}{\beta_{m,1}}e^{jk_{t_{m,1}}y}.
\end{aligned}
\end{equation}
The first term corresponds to the source fields that the grid currents induce at the reference strip position (in the absence of the substrate and PEC); it originates from \eqref{eq:field_from_wire_source} after applying the Poisson formula and taking $\delta\rightarrow0$. However, the Hankel function diverges at $(y,z)=(0,-h)$ and therefore this transformation is not applicable \textcolor{black}{at this point}. 

\textcolor{black}{Hence, we retrace our steps and utilize  the original form \eqref{eq:field_from_wire_source} for the field assessment. This is achieved by following \cite{tretyakov2003analytical}, separating the Hankel function summation} into the self-induced fields that the reference wire creates on its shell \textcolor{black}{($y\rightarrow r_\mathrm{eff}$)}, and the field generated by all the other strips at the position of the reference strip\textcolor{black}{, reading
\begin{equation}
\label{eq:E_wire_air_delta_0_Hankel_Appendix}
\begin{aligned}
&E^\mathrm{grid}_{x,1}(y\rightarrow0,z\rightarrow-h)=-\frac{k_{1}\eta_{1}}{4}IH^{(2)}_{0}(k_{1}r_\mathrm{eff})\\
&-\!\!\frac{k_{1}\eta_{1}}{4}I\!\!\!\!\sum_{\substack{n=-\infty \\ n\neq 0}}^{\infty}\!\!\!\!e^{-jk_{1}n\Lambda\sin \theta_\mathrm{in}}
H^{(2)}_{0}(k_{1}\left|n\Lambda\right|)\\
&-\frac{k_{1}\eta_{1}}{2\Lambda}I\!\!\!\!\sum_{m=-\infty}^{\infty}\frac{R_{m}}{\beta_{m,1}}.
\end{aligned}
\end{equation}
}

\textcolor{black}{As in \cite{tretyakov2003analytical,epstein2017unveiling}, the first term in \eqref{eq:E_wire_air_delta_0_Hankel_Appendix} can be approximated by the asymptotic expression of the Hankel function for small arguments as per Eq. (9.1.8) in \cite{AbramowitzStegun1970} (recall that $w=4r_\mathrm{eff}\ll\lambda$), while the second term can be expanded using Eq. (8.522) in \cite{GradshteinRyzhik2015}. The last term corresponds to the field induced on the reference strip due to the reflections of the grid-generated field from the PEC-backed dielectric; it can be interpreted as the field due to multiple image sources positioned at $z>0$ \cite{felsen1994radiation}, thus does not suffer of convergence issues at the distant reference strip, and can be evaluated directly. Applying these transformations onto \eqref{eq:E_wire_air_delta_0_Hankel_Appendix} and rearranging the terms leads to (\ref{eq:equation_on_load}).}
%

\bibliographystyle{IEEEtran}
\bibliography{AnalyticalPCBMetagratings}

\end{document}